\documentstyle[12pt,oldlfont]{article}
\begin{document}

\newcommand{\non}{\nonumber}
\newcommand{\nn}{\noindent}
\newcommand{\s}{\\ \vspace*{-2mm}}
\newcommand{\nt}{\not \hspace*{-1mm} }
\newcommand{\be}{\begin{eqnarray}}
\newcommand{\en}{\end{eqnarray}}
\newcommand{\sla}{\lambda^{\frac{1}{2}}}
\newcommand{\lam}{\lambda^{\frac{1}{2}}}

\renewcommand{\thefootnote}{\fnsymbol{footnote} }

\nn \hspace*{12cm} UdeM-LPN-TH-93-169 \\
    \hspace*{12cm} NYU--TH--93/09/04 \\
\hspace*{12cm} September 1993 \\

\vspace*{1cm}

\centerline{\large{\bf Electroweak gauge bosons self--energies: }}

\vspace*{0.4cm}

\centerline{\large{\bf complete QCD corrections.} }

\vspace*{1.5cm}

\centerline{\sc A.~Djouadi$^1\footnote{NSERC Fellow.}$ and P.~Gambino$^2$.}

\vspace*{1cm}

\centerline{$^1$ Laboratoire de Physique Nucl\'eaire, Universit\'e de
Montr\'eal,  Case 6128 Suc.~A,}
\centerline{H3C 3J7 Montr\'eal PQ, Canada.}

\vspace*{0.4cm}

\centerline{$^2$ Department of Physics, New York University, 4 Washington
Place,}
\centerline{ New York, NY 10003, USA.}

\vspace*{2cm}

\begin{center}
\parbox{14cm}
{\begin{center} ABSTRACT \end{center}
\vspace*{0.2cm}

\nn We present the QCD corrections to the longitudinal and transverse
components
of the electroweak gauge bosons self--energies for arbitrary momentum transfer
and for different internal quark masses. Compact formulae for both the real and
imaginary parts are given in the general case as well as in some physically
interesting special cases. The dependence on the definition of the quark masses
is discussed. }

\end{center}

\newpage

\renewcommand{\thefootnote}{\arabic{footnote} }
\setcounter{footnote}{0}

\subsection*{1.~Introduction}

The Standard Model of the electroweak and strong interactions has achieved a
tremendous success in describing all experimental data within the range of
energies available today. In particular, it has been tested to the level of its
quantum corrections in the high--precision LEP and SLC experiments \cite{lep}.
To allow for such precision tests, a considerable amount of theoretical work
has been done in the last few years to calculate the relevant radiative
corrections \cite{zp}. In fact, in view of the remarkably high accuracy of the
experimental data, some radiative corrections were required beyond the
one--loop approximation. \s

\nn Except for the QCD corrections to light quark pair production in $e^+e^-$
annihilation, which are known up to ${\cal O}(\alpha_S^2)$ \cite{two} since
more than a decade and which have been recently calculated to ${\cal O}(
\alpha_S^3)$ \cite{three}, these two--loop corrections were mainly concerned
with the effects of the top--bottom weak isodoublet. Indeed, besides the latter
QCD corrections and the well--known electromagnetic corrections \cite{zp},
these genuine electroweak effects are by far the most sizeable: top quark
induced radiative corrections appear in the self--energies of the weak vector
bosons \cite{vel} and in the coupling of the $Z$ boson to bottom quarks
\cite{Zbb} and lead to contributions which are quadratically proportional to
the large mass \cite{mt} of this yet unobserved particle. Two types
of higher order corrections to the previous quantities have been calculated so
far. The first one involves virtual Higgs boson exchange which leads to
potentially large ${\cal O}(\alpha^2 m_t^4/M_W^4)$ contributions \cite{higgs}.
Motivated by the large value of the QCD coupling constant, the second type of
corrections are the two--loop ${\cal O} (\alpha \alpha_S)$ contributions which,
for the transverse components of the vector bosons self--energies have been
calculated in Ref.~[9--13] and extensively discussed in Ref.~[14--16], and for
the $Z$ boson coupling to bottom quarks derived recently in Ref.~\cite{vertex}.
\s

\nn However, due to the great technical difficulties that one encounters when
calculating at this level of perturbation theory, all the previously mentioned
results involve some simplifying approximations. For instance, the Higgs
exchange contributions to both the weak bosons self--energies and to the
$Z$--bottom quarks vertex as well as the QCD corrections to the latter vertex,
were calculated in the limit where the top quark is much heavier than the weak
vector bosons, an approximation which might turn out to be very poor in view of
the
lower limit on the top quark mass \cite{mt}. A better approximation has been
made in the case of the QCD corrections to the vector bosons self--energies
since results are available in the limit of a vanishing bottom quark mass; this
is certainly a good enough approximation for practical purposes since the
bottom quark mass amounts to, at most, a few percent of the top quark mass.
\s

\nn From a theoretical viewpoint, however, it would be desirable to have a
result with no approximation involved except, of course, for the one due to the
truncation of the perturbative series. This would place analytical calculations
in the Standard Model at a level comparable to what is known in QED where, in a
pioneering work, the authors of Ref.~\cite{QED} have derived exactly the vacuum
polarization function of the photon at two--loop order. In addition, an exact
calculation might turn out to be mandatory in the case of a fourth generation,
the existence of which is still allowed by present experimental data if the
associated neutrino is heavy enough \cite{lep}. \s

\nn In this article, we present exact and compact analytical expressions for
the ${\cal O}(\alpha \alpha_S)$ contributions of quark pairs to the vacuum
polarization functions of the electroweak gauge bosons in the most general
case:
real and imaginary parts of both the transverse and longitudinal components,
for different and non--zero quark masses and for arbitrary momentum transfer.
\s

\nn The paper is organized as follows. In the next section we summarize all the
one--loop results which will be relevant to our discussion. In section 3, we
discuss the dependence of the result on the definition of the quark masses and
provide the material necessary to derive the vector bosons self--energies in
any renormalization scheme. The expressions for the real and imaginary parts of
the vacuum polarization function in the most general case are given in section
4. In section 5, we display the expressions of the self--energies in special
situations of physical interest: equal quark masses, one massless quark and at
zero or very high momentum transfer. Finally, section 6 contains our
conclusions. In the Appendix, we will list for completeness the expressions of
the one and two--loop scalar integrals that we encountered in this calculation.

\subsection*{2.~One--loop results}

\renewcommand{\theequation}{2.\arabic{equation}}
\setcounter{equation}{0}

\nn To set the notation and for the sake of completeness, we rederive in this
section all the one--loop results which will be relevant to our discussion. \\

\nn The contribution of a fermionic loop to the vacuum polarization tensor of a
vector boson $i$, or to the mixing amplitude of two bosons $i$ and $j$, denoted
$\Pi_{\mu\nu}^{ij}$, is defined as
\be
\Pi_{\mu\nu}^{ij}(q^2)= -i \int d^4 x e^{i q\cdot x}<0|{\rm T}^* \left[J_\mu^{
i}(x) J_\nu^{j \dagger }(0) \right]|0>
\label{def} \en
where T$^*$ is the covariant time ordering product and $q$ the four--momentum
transfer; $J_\mu^{i}, J_\nu^{j}$ are fermionic currents coupled to the vector
bosons $i,j$ and constructed with spinor fields whose corresponding masses are
$m_a, m_b$. The definition of $\Pi_{\mu\nu}^{ij}$ corresponds to $+i$ times the
standard Feynman amplitude. The vacuum polarization tensor can be decomposed
into a transverse and a longitudinal part,
\be
\Pi_{\mu\nu}^{ij}(q^2)=\left(g_{\mu \nu}-\frac{q_\mu q_\nu}{q^2}
\right) \Pi_T^{ij}(q^2)\ + \ \frac{q_\mu q_\nu}{q^2}\ \Pi_L^{ij}(q^2)
\en
and the two components can be directly extracted by contracting $\Pi_{\mu \nu
}^{ij} (q^2)$ by the two projectors $g_{\mu \nu}-q_\mu q_\nu/q^2$ and
$q_\mu q_\nu/q^2$; in $n=4-2\epsilon$ dimensions one has
\be
\Pi^{ij}_T (q^2) & = & \frac{1}{3-2\epsilon} \left(g^{\mu \nu}-\frac{q^\mu
q^\nu} {q^2} \right) \Pi^{ij}_{\mu \nu} (q^2) \nn \\
\Pi_{L}^{ij}(q^2)& = & \frac{q^\mu q^\nu}{q^2} \Pi^{ij}_{\mu \nu} (q^2)
\en
\nn The one--loop transverse and longitudinal components can be written as
[with $s \equiv q^2$]
\be
\Pi_{T,L}^{ij}(s)= \frac{\alpha}{\pi} N_C \left[ \ (v^iv^j+a^ia^j) s \
\Pi_{T,L}^+ (s) \ + \ (v^iv^j-a^ia^j) \ m_a m_b \ \Pi_{T,L}^- (s)\
\right]
\en
with $N_C$ the number of colors and $v^i$ and $a^i$ the vector and
axial--vector
couplings of the gauge boson $i$ to the fermions expressed in units of the
proton charge $e=\sqrt{4 \pi \alpha}$. \s

\nn The vector and axial--vector components of the vacuum polarization function
[with the coupling constants factored out] are then simply given by
\be
\Pi^{V,A}_{T,L} (s) = s \Pi^+_{T,L} (s) \ \pm \ m_a m_b \Pi^-_{T,L}(s)
\en
which exhibits the fact that $\Pi^{A,V}(s)$ can be obtained from $\Pi^{V,A}(s)$
by simply making the substitution $m_a(m_b) \rightarrow -m_a(-m_b$) as
expected from $\gamma_5$ reflection symmetry. \s

\nn At the one--loop level and for arbitrary fermion masses, $m_a
\neq m_b \neq 0$, the vacuum polarization amplitude, Fig.~1a, reads
\be
\Pi_{\mu \nu}^{ij}(q^2)= - i e^2  N_C \left( \frac{\mu^2 e^\gamma}{4 \pi}
\right)^\epsilon \int \frac{d^n k}{(2\pi)^n} {\rm Tr} \ \frac{(\nt k + m_a)
\gamma_\mu (v^i - a^i\gamma_5) (\nt k - \nt q + m_b) \gamma_\nu (v^j-a^j
\gamma_5)} {(k^2-m_a^2) [(k-q)^2-m_b^2]}
\en
where $\mu$ is the 't Hooft renormalization mass scale introduced to make the
coupling constant dimensionless in $n=4-2\epsilon$ dimensions; we have also
introduced an extra term $(e^\gamma /4\pi)^\epsilon$ [$\gamma$ is the Euler
constant] to prevent uninteresting combinations of $\log4\pi, \ \gamma \cdots$,
in the final result. After contracting the amplitude with the tensors $(g^{\mu
\nu}-q^\mu q^\nu /q^2)$ and $q^\mu q^\nu/q^2$, and expressing the scalar
products of the momenta appearing in the numerator in terms of combinations of
the polynomials appearing in the denominators, one is led to the calculation of
two scalar one--loop integrals which, for completeness, are given in Appendix.
\s

\nn The expressions of $\Pi^{\pm}_{T,L}(s)$ in the general case $m_a \neq m_b
\neq 0$ are then
\be
\Pi_T^+ (s) &=& \frac{2+3\alpha+3 \beta}{6 \epsilon} -\frac{1}{4} \left(
\frac{2}{3} +\alpha + \beta \right) (\rho_a +\rho_b)  + \frac{5}{9}+\frac{1}{3}
(\alpha+\beta) - \frac{1}{6} (\alpha - \beta)^2 \non \\
& -& \ \frac{1}{12} (\alpha- \beta)^3 \log \frac{\alpha}{\beta} + \frac{1}{12}
\sla \left[ 3(1+\alpha+ \beta)-\lambda \right] (\log x_a + \log x_b ) \non \\
\Pi_L^+ (s) &=& \frac{\alpha+\beta}{2 \epsilon} -\frac{1}{4} (\alpha+
\beta) (\rho_a +\rho_b) +  \alpha+ \beta + \frac{1}{2} (\alpha - \beta)^2
+\frac{1}{4} (\lambda -1) (\alpha -\beta) \log \frac{\alpha}{\beta} \non \\
& + & \ \frac{1}{4} \sla (\lambda-1-\alpha  -\beta) (\log x_a + \log x_b ) \non
\\
\Pi_{T}^-(s) &=& \Pi_{L}^-(s) = \frac{1}{\epsilon} - \frac{1}{2} (\rho_a
+\rho_b) + 2 + \frac{1}{2} (\alpha-\beta) \log \frac{\alpha}{\beta}
+ \frac{1}{2} \sla (\log x_a + \log x_b )
\en
where we use the variables [~$\beta, \rho_b$ and $x_b$ are defined similarly]
\be
\alpha =-{m_{a}^2\over s} \ \ , \ \ \rho_{a}=\log{m_{a}^2\over \mu^2} \ \ ,
\ \ \ \ x_a =\frac{2 \alpha}{1+\alpha+\beta+\sla }
\en
\be
\lambda= 1+2\alpha +2\beta+ (\alpha -\beta)^2
\en

\vspace*{0.3cm}

\nn From these general formulae, it is straightforward to derive the
expressions
of the self--energies in two special situations of physical interest: $i)$ the
two quarks have equal masses, which corresponds to the case of a neutral gauge
boson [the photon or the $Z$ boson] and $ii)$ one of the quarks has zero--mass,
which corresponds to the case of a charged gauge boson with one quark much
heavier than the other [as for the top--bottom isodoublet]. For completeness,
we exhibit in the following the expressions of the vector and axial--vector
parts of the transverse and longitudinal\footnote{Note that the contribution
of the longitudinal part in processes involving fermions in the initial and/or
final states is always suppressed by the masses of these fermions.} components
of the gauge bosons self--energies in these two limiting situations. \\

\nn For $m_b=m_a$, one has
\be
\Pi^V_T (s) &=& \frac{s}{3} \left[ \frac{1}{\epsilon}-\rho_a +\frac{5}{3}
-4 \alpha + (1+4\alpha)^{1/2} (1- 2\alpha)\log \frac{4 \alpha}{(1+
\sqrt{1+4\alpha})^2} \right] \non \\
\Pi^A_T (s) &=& \frac{s}{3} \left[ \left( \frac{1}{\epsilon}-\rho_a \right)
(1+6\alpha) +\frac{5}{3}+ 8 \alpha + (1+4\alpha)^{3/2} \log
\frac{4 \alpha}{(1+\sqrt{1+4\alpha})^2} \right] \non \\
\Pi^V_L (s) &=& 0 \non \\
\Pi^A_L (s) &=& 2s \alpha \left[ \frac{1}{\epsilon}-\rho_a +2+(1+4\alpha)^{1/2}
 \log \frac{4\alpha}{(1+\sqrt{1+4\alpha})^2}
\right]
\en

\vspace*{0.1cm}

\nn while for $m_b=0$, one has
\be
\Pi_T^{V,A} (s) & =& \frac{s}{6} \left[ (2+3\alpha) \left( \frac{1}
{\epsilon}-\rho_a \right) + \frac{10}{3} + 2 \alpha - \alpha^2 + (2+3\alpha
-\alpha^3 ) \log \frac{\alpha}{1+\alpha} \right] \non \\
\Pi_L^{V,A} (s) & = & \frac{s}{2} \alpha \left[ \frac{1}{\epsilon}
-\rho_a +2 + \alpha + (1+\alpha)^2 \log \frac{\alpha}{1+\alpha} \right]
\en

\vspace*{0.4cm}

\nn In many cases, it is of interest to evaluate the vacuum polarization
functions at zero momentum transfer. This is the case for instance of the
radiative correction to the $\rho$ parameter \cite{vel} which measures the
relative strength of the charged to neutral current at $q^2=0$. In this limit,
the contributions of a fermionic loop to the transverse and longitudinal
components simplify to
\be
\Pi^{V,A}_{T,L} (0) &=& - \frac{1}{2\epsilon} (m_a^2+m_b^2)
-\frac{1}{4}(m_a^2+m_b^2)+ \frac{1}{4}(m_a^2+m_b^2)(\rho_a+\rho_b) +\frac{1}{4}
\frac{m_a^4+m_b^4}{m_a^2-m_b^2} \log \frac{m_a^2}{m_b^2} \non \\
&& \pm \ m_a m_b \left[ \frac{1}{\epsilon} +1 - \frac{1}{2}(\rho_a
+\rho_b) -\frac{1}{2} \frac{m_a^2+m_b^2}{m_a^2-m_b^2} \log \frac{m_a^2}{m_b^2}
\right]
\en

\nn Note that in this limit, the longitudinal and transverse components of the
vacuum polarization function are equal and that for a vector current with equal
fermion masses $m_a= m_b$ [as it is the case for the photon] the vacuum
polarization function vanishes. \\

\nn In the opposite limit $m_a^2, m_b^2 \ll s$, the vector and axial--vector
components of the gauge bosons self--energies simply reduce to
\be
\Pi_T^{V,A} (s) & \sim & \frac{s}{3} \left[ \frac{1}{\epsilon} -\log \frac{-s}
{\mu^2} +\frac{5}{3} \right] \non  \\
\Pi_L^{V,A} (s) & \sim & 0
\en
\nn One sees that $\Pi_{T,L}^{V}(s)=\Pi_{T,L}^{A}(s)$ as expected from chiral
symmetry and that, in this limit, there is no mass singularity. Note that
$\Pi_{T}^{V,A}(s)$ are quadratically divergent for $ |q| \to \infty$. \\

\vspace*{3mm}

\nn In all the previous expressions the momentum transfer has been defined to
be in the space--like region, $s<0$. When continued to the physical region
above the threshold for the production of two fermions, $s \geq
(m_a+m_b)^2$, the vector boson self--energies acquire imaginary parts, the
latter being related to the decay widths of the vector bosons into fermions.
Adding a small imaginary part $-i \epsilon$ to the fermion masses squared, the
analytical continuation is consistently defined. \s

\nn From the expressions eqs.~(2.8), the imaginary parts can be
straightforwardly obtained by making the substitution
\be
\log x_{a,b} \longrightarrow \log|x_{a,b}| + i \pi \non
\en
and one has
\be
{\cal I}m \Pi_T^+(s) = \frac{\pi}{2} \sla \left[ (1+\alpha+\beta)- \frac{1}{3}
\lambda \right]  \hspace*{0.7cm} , \hspace*{0.7cm} {\cal I}m \Pi^-_T (s) = \pi
\sla  \non \\
{\cal I}m\Pi^+_L (s) = \frac{\pi}{2} \sla \left[ \lambda -1 - \alpha-\beta
\right] \hspace*{1cm} , \hspace*{0.7cm} {\cal I}m\Pi^-_L (s) = \pi \sla
\en
or, equivalently,
\be
{\cal I}m \Pi_T^{V,A} (s) &=& \frac{\pi }{2} s \sla \left[ (1+\alpha+\beta)-
\frac{1}{3} \lambda \ \pm \ 2 \frac{m_a m_b}{s} \ \right]  \non \\
{\cal I}m \Pi_L^{V,A} (s) &=& \frac{\pi}{2} s \sla \left[ \lambda -1 - \alpha-
\beta \ \pm \ 2 \frac{m_a m_b}{s} \right]
\en
\nn The knowledge of the imaginary part of the polarization function, which can
be calculated directly using Cutkosky rules \cite{cut}, allows an alternative
way for obtaining its real part $\Pi(s)$: the latter can be expressed as a
dispersive integral of ${\cal I}m \Pi(s)$. The connection between the
dispersive approach and the results which are derived using dimensional
regularization, has been discussed in Ref.~\cite{ks} in the special cases
$m_a=m_b$ and $m_b=0$ and more recently \cite{nous} in the general case.

\newpage

\subsection*{3.~Renormalization and scheme dependence}

\renewcommand{\theequation}{3.\arabic{equation}}
\setcounter{equation}{0}

\nn At ${\cal O}(\alpha \alpha_S)$, the two--loop diagrams contributing to the
vacuum polarization function $\Pi_{\mu \nu}^{ij}(q^2)$ induced by quark loops
[up to a factor $+i$] are shown in Fig.~1b. In the 't Hooft--Feynman gauge,
using the routing of momenta shown in the figure and following the notations
introduced in the previous section, one can write the bare amplitude as
\be
\left. \Pi_{\mu \nu}^{ij}(q^2) \right|_{\rm bare}= -\frac{4}{3} \alpha \alpha_S
(16 \pi^2) \ \left( \frac{\mu^2 e^\gamma}{4\pi} \right)^{2\epsilon} \ \int
\frac{d^nk_1}{(2\pi)^n} \ \int \frac{d^n k_2}{(2\pi)^n} \ \left[ {\cal A}_{\mu
\nu}^{ij} +{\cal B}_{\mu \nu}^{ij} \right]
\en
with
\be
{\cal A}_{\mu \nu}^{ij} &=& {\rm Tr} \frac{ (\nt k_1+m_a)\gamma_{\mu}(v^i-a^i
\gamma_5) (\nt k_1 -\nt q +m_b) \gamma_\lambda (\nt k_2 -\nt q +m_b) \gamma_\nu
(v^j-a^j \gamma^5) (\nt k_2+m_a) \gamma^\lambda }{ (k_1-k_2)^2 \ (k_1^2-m_a^2)
\ (k_2^2-m_a^2) \ [(k_1-q)^2-m_b^2] \ [(k_2-q)^2-m_b^2]} \non \\
{\cal B}_{\mu \nu}^{ij} &=& {\rm Tr} \frac{ (\nt k_1+m_a)\gamma_{\mu}(v^i-a^i
\gamma_5) (\nt k_1 -\nt q +m_b) \gamma_\nu (v^j-a^j \gamma^5) (\nt k_1+m_a)
\gamma_\lambda (\nt k_2 +m_a) \gamma^\lambda}{ (k_1-k_2)^2 \
(k_1^2-m_a^2)^2 \ (k_2^2-m_a^2) \ [(k_1-q)^2-m_b^2] } \non \\
& &   \ + \ m_a \ \longleftrightarrow \ m_b
\en

\nn This bare amplitude has to be supplemented by counterterms. By virtue of
the QED--like Ward identity, the vertex and fermion wave function counterterms
cancel each other and only quark mass renormalization has to be included. The
latter is obtained by considering the diagram shown in Fig.~2a, the amplitude
of
which reads in dimensional regularization
\be
-i \Sigma(\nt p) = - \alpha_s \frac{16 \pi}{3} \left( \frac{e^\gamma \mu^2 }
{4\pi} \right)^\epsilon \int \frac{d^nk}{(2\pi)^n} \frac{ \gamma^\lambda (\nt p
- \nt k + m) \gamma_\lambda } {[(p-k)^2-m^2] \ k^2}
\en
where $p$ is the four--momentum of the quark and $m$ its bare mass. This
expression can be decomposed into a piece proportional to $(\nt p-m)$ which
will enter the wave function renormalization and another piece proportional
to $m$ which will give the mass counterterm. After integration over the loop
momentum, the latter is given by
\be
\Sigma_m (p^2) =\frac{\alpha_S}{\pi} \left( \frac{\mu^2 e^\gamma}{m^2} \right)
^\epsilon \frac{m}{\epsilon} \ \frac{1-2\epsilon/3}{1-2\epsilon} \Gamma(1+
\epsilon) \left[ 1+ \left(\frac{p^2}{m^2} -1 \right) {\cal O}(\epsilon) \right]
\en
The mass counterterm will now depend on the renormalization procedure; i.e. on
the definition of the quark mass. For instance, in the on--shell scheme which
is usually used to calculate radiative corrections in the electroweak theory
\cite{sir}, the fermion masses are defined at $p^2=m^2$ and correspond to the
position of the pole of the fermion propagators. They are referred to as the
on--shell or physical masses and the counterterms will be given by
\be
\delta m \ \equiv  \ m (m^2) - m \ = \ \frac{\alpha_S}{\pi} \frac{m }{\epsilon}
\left( \frac{\mu^2}{m^2}\right)^\epsilon \ \left( 1+ \frac{\pi^2}{12}\epsilon^2
\right) \ \frac{1-2\epsilon/3}{1-2\epsilon}
\en
One then inserts this mass counterterm in the one--loop self--energies, as
depicted in the diagrams of Fig.~2b, which is equivalent to calculate
\be
\left. \Pi_{\mu \nu}^{ij}(q^2) \right|_{\rm CT} = \ - \delta m_a \frac{
\partial}{\partial m_a} \left. \Pi_{\mu \nu}^{ij} (q^2) \right|_{\rm 1-loop}
\ - \delta m_b \frac{\partial}{\partial m_b } \left. \Pi_{\mu \nu}^{ij}
(q^2)\right|_{\rm 1-loop}
\en
where the one--loop vacuum polarization function is given by eqs.~(2.7--2.8).
The renormalized two--loop self--energies will then read
\be
\Pi_{\mu \nu}^{ij} (q^2) = \left. \Pi_{\mu \nu}^{ij} (q^2) \right|_{\rm bare}
+\left. \Pi_{\mu \nu}^{ij} (q^2) \right|_{\rm CT}
\en

\vspace*{0.3cm}

\nn However, one can also employ a different definition for the quark mass. For
instance one can use the $\overline{\rm MS}$ scheme in which the mass is
defined
by just picking the divergent term in the expression of $\Sigma_m(p^2)$ [the
related constants $\log 4\pi, \gamma, \cdots$ have already been subtracted] or
the running mass\footnote{A discussion of the phenomenological consequences of
using a different definition for the quark masses will appear elsewhere
\cite{let}.}.
Having at hand the result of the vacuum polarization function in the on--shell
scheme, one can obtain the polarization function in any other scheme X: one
simply has to add to the expression of the two--loop self--energy eq.~(3.7) the
quantity\footnote{Note that since at this stage we are only discussing the
difference between two renormalization schemes, we do not need the ${\cal O}
(\epsilon)$ terms in the one--loop result of the vector bosons
self--energies. These terms are of course needed to evaluate the renormalized
two--loop self--energies in a given scheme.}
\be
\left. \Delta \Pi_{\mu \nu}^{ij} (q^2) \right|_X = \Delta m_a^X  \frac{m_a
\partial}{\partial m_a}\left. \Pi_{\mu \nu}^{ij} (q^2) \right|_{\rm 1-loop} +
\Delta m_b^X \frac{m_b \partial}{\partial m_b } \left. \Pi_{\mu \nu}^{ij}
(q^2)\right|_{\rm 1-loop}
\en
where in terms of the quark mass $m_{a,b}^X$ defined in the scheme $X$, one has
\be
\Delta m_{a,b}^X = \left[ 1- \frac{m_{a,b}^X} {m_{a,b}(m_{a,b}^2)} \right]
\en

\nn Decomposing the one--loop function into its transverse and longitudinal
components as in eq.~(2.5), one obtains for the derivative of these components
\be
m_a \frac{\partial}{\partial m_a} \left. \Pi_T^+ (s) \right|_{\rm 1-loop}
&=& \frac{\alpha}{\epsilon} - \frac{1}{2} \alpha(\rho_a +\rho_b) + \alpha
(\beta-\alpha)- \frac{1}{2} \alpha(\alpha-\beta)^2 \log \frac{\alpha}{\beta}
\non \\
&+&\frac{1}{2} \lam \ \alpha( 1-\alpha+\beta)(1 -2\beta \lambda^{-1} )(\log x_a
+ \log x_b ) \non \\
m_a \frac{\partial}{\partial m_a} \left. \Pi_L^+ (s) \right|_{\rm 1-loop}
&=& \frac{\alpha}{\epsilon} - \frac{1}{2} \alpha(\rho_a +\rho_b) + \alpha
(2-3\beta+3\alpha)+ \frac{1}{2} \alpha [3(\alpha-\beta)^2+4\alpha ] \log
\frac{\alpha}{\beta} \non \\
&+& \frac{1}{2} \lam \ \alpha \left[1+3\alpha-3\beta +2\beta \lambda^{-1} (1+
\beta-\alpha) \right] (\log x_a + \log x_b ) \non
\en
\be
m_a \frac{\partial}{\partial m_a} \left. m_a m_b \Pi_{T,L}^- (s) \right|_{\rm
1-loop} &=& m_a m_b \left[ \frac{1}{\epsilon} - \frac{1}{2} (\rho_a +\rho_b) +
2 +\frac{1}{2} (3\alpha-\beta) \log \frac{\alpha}{\beta} \right. \non \\
&+& \left. \frac{1}{2} \lam \left[1+ 2\alpha \lambda^{-1} (1+\alpha-\beta)
\right] (\log x_a + \log x_b ) \right] \hspace*{1.5cm}
\en
and similarly for the piece involving $m_b$ which can be obtained by making
the substitution $m_a \leftrightarrow m_b$. Hence, once the complete result
of the vacuum polarization function is known in the on--shell scheme, it is
straightforward using eqs.~(3.8)--(3.10), to obtain the corresponding
results in any renormalization scheme. At this stage, a few remarks are
mandatory. \s

\nn First, one notices in the previous expressions the occurence of terms that
are inversely proportional to the velocity factor $\lambda^{1/2}$. In
principle, after an analytical continuation to the physical region beyond the
threshold for quark pair production, $s \geq (m_a+m_b)^2$, these terms would
diverge for energy values near the production threshold, $\lambda \sim 0$.
However, as we will see later, when evaluated in the on--shell scheme, which is
the only scheme where the physical threshold is well defined, the renormalized
two--loop polarization function is free of these $\lambda^{-1/2}$ factors. Near
threshold, the dominant terms will be constants and would correspond, once the
vector boson self--energy is \underline{normalized} to its one--loop value, to
the well known $\lambda^{-1/2}$ Coulomb singularities which require a
non--pertubative treatment; see for example Ref.~\cite{ks}. Of course, the
$\lambda^{-1/2}$ terms can be present in the vacuum polarization function when
it is evaluated in a different scheme but in this case, the threshold is not
well defined since the masses are not ``physical" masses. \s

\nn In principle one can define the masses $m_a$ and $m_b$ in two completely
different schemes; this can be useful if, for instance, different scales are
involved in the evaluation of the vacuum polarization functions. Taking the
example of the top--bottom isodoublet, one can employ the on--shell mass for
the heavy top quark, which is suitable if one wants to discuss threshold and
possibly non--pertubative effects \cite{ks}; and use a running mass evaluated
at the scale $q^2$ for the relatively light $b$ quark, which in general, avoids
the appearence of large logarithms for $q^2 \gg m_b^2$ \cite{run}. In the rest
of our discussion, however, we will stick to the on--shell mass scheme. \s

\nn Finally, the scale at which the strong coupling constant $\alpha_S$ is
evaluated can, of course, be completely different from the one chosen for the
quark masses. In general, $\alpha_S$ is evaluated at the scale of the problem
at hand, i.e. $\alpha_S \equiv \alpha_S(s)$. However, for heavy virtual quarks,
the effective coupling is usually taken at the mass of the quark \cite{SVZ}
and this choice can be justified by arguments based on effective field theory
\cite{SSCL} [for a discussion on this point see also the second paper of
Ref.~\cite{ks}]. For instance, in the case of the top--bottom isodoublet, the
two--loop QCD corrections to electroweak observables are calculated with
$\alpha_S(m_t^2)$ while for light quarks, one uses\footnote{An exception is
the contribution of the light quarks to the photon self--energy which often has
to be evaluated at zero momentum transfer. Since both $q^2$ and $m^2$ are small
compared to the QCD scale, one can no more use perturbation theory. However,
the problem can be circumvented by using a dispersive approach and relating the
photon self--energy to $e^+ e^- \rightarrow$ hadrons low energy data
\cite{BJV}.} $\alpha_S(q^2)$.

\subsection*{4.~Exact two--loop results}
\renewcommand{\theequation}{4.\arabic{equation}}
\setcounter{equation}{0}

In this section we give the expression of the vacuum polarization function at
order ${\cal O}(\alpha \alpha_S)$ in the general case $m_a \neq m_b \neq 0$
and for arbitrary momentum transfer. The result will be given in the on--shell
mass scheme. We will follow very closely the notations and definitions
introduced in section 2; since confusion should be rare, we will use the same
notation for the one--loop and two--loop self--energies and $m_{a,b}$ will
stand for the on--shell masses. \\

\nn At ${\cal O}(\alpha \alpha_S)$, the transverse and longitudinal components
of the vacuum polarization function $\Pi_{\mu \nu}^{ij}(q^2)$ are defined by
[the color factor $N_C$ is now included]
\be
\Pi_{T,L}^{ij}(s)= \frac{\alpha}{\pi} \frac{\alpha_S}{\pi} \left[ \ (v^iv^j+
a^ia^j) s \ \Pi_{T,L}^+ (s) \ + \ (v^iv^j-a^ia^j) \ m_a m_b \ \Pi_{T,L}^- (s)\
\right]
\en
where $\Pi_{T,L}^\pm$ are the sum of the corresponding components in the bare
two--loop amplitude eqs.~(3.1--3.2) and the mass counterterm which can be
obtained from eq.~(3.11). \s

\nn Similarly to the one--loop case, after contracting $\Pi_{\mu \nu}^{ij}$ in
eqs.~(3.1) by the tensors $(g^{\mu \nu}-q^\mu q^\nu/q^2)$ and $q^\mu q^\nu/q^2$
and expressing the scalar products of momenta appearing in the numerators in
terms of combinations of the polynomials in the denominators, one is led to the
calculation of a set of scalar two--loop integrals. Most of these integrals
have
been first calculated by Broadhurst in Ref.~\cite{bro}; the remaining integrals
reduce after straightforward computations to the previous ones \cite{gen,moi2}.
For completeness, these scalar integrals will be given in the Appendix. \s

\nn After a cumbersome computation, and taking advantage of the symmetry
in the change $\alpha \leftrightarrow \beta$, we can write $\Pi^\pm_{T,L}(s)$
in a relatively simple and compact form
\be
\Pi_T^+ &=& \left\{ - \frac{3\alpha }{2\epsilon^2}+\frac{1}
{\epsilon} \left( \frac{1}{4}+ 3 \alpha \rho_a - \frac{11}{4} \alpha \right)
+ \frac{1}{4} (\rho_a+\rho_b) \left(11 \alpha-1-9\alpha \rho_a+3\alpha \rho_b
\right) + \frac{55}{24} -\frac{71}{24} \alpha \right. \non \\
&- & \frac{5}{6} \alpha^2 +\frac{11}
{6} \alpha \beta +\frac{2}{3}\alpha [G(x_b)-G(x_a)]+ \frac{1}{4} \log x_a
\log x_b \left[ 3(\alpha+\beta)+2\alpha \beta(4+\alpha+\beta) \right] \non \\
& + & \frac{1}{12} \log x_a \left[ \left( \alpha-\beta + \lam
\right)(11+19\alpha+19 \beta + 12\alpha\beta-5 \lambda)+(\alpha-\beta)(42-5
\alpha -5\beta) \right] \non \\
&+ & \frac{1}{12} \log ^2 x_a \left[ (1-3\alpha-3\beta)\left( \lambda-1-\alpha-
\beta +(\alpha-\beta) \lam \right) -9(\alpha+\beta)+8\alpha
\beta \right] - \frac{\pi^2}{4} \alpha  \non \\
& + & \left. \frac{1}{6} \left[(\alpha+\beta-2) \lambda -12 \alpha \beta
\right]
{\cal I} -\frac{1}{3} \left[ 3(1+\alpha+\beta)- \lambda \right] {\cal I}'
\right\} \ \ + \ \ \left\{ \alpha \longleftrightarrow \beta \right\} \non
\en
\be
\Pi_T^- &=& \left\{ - \frac{3}{2\epsilon^2} +\frac{1}{\epsilon}
\left(3\rho_a - \frac{11}{4}\right) + \frac{11}{2}\rho_a -\frac{3}{4} (\rho_a+
\rho_b)^2 +\frac{11}{8} + \alpha - \frac{\pi^2}{4} \right. \non \\
&+ & \frac{1}{2} \log x_a \left[ \alpha-\beta + (\alpha+\beta+9)(\alpha-\beta+
\lam ) \right] + \frac{1}{2}\log ^2 x_a \left[ 1+\alpha+\beta-
\lambda + (\beta- \alpha) \lam \right] \non \\
&+ & \left. \frac{1}{2} \log x_a \log x_b \left[ 3(1+\alpha+\beta)+2\alpha
\beta
\right] -(1+\alpha+\beta){\cal I} -2{\cal I}' \right\} \ \ + \ \
\left\{ \alpha \longleftrightarrow \beta \right\} \non
\en
\be
\Pi_L^+ &=& \left\{ - \frac{3\alpha }{2\epsilon^2}+\frac{1}{\epsilon} \left( 3
\alpha \rho_a - \frac{11}{4} \alpha \right) + \frac{\alpha}{4} (\rho_a+\rho_b)
\left(11 -9\rho_a+3\rho_b \right) + \frac{3}{8} \alpha + \frac{7}{2} \alpha^2
-\frac{13}{2} \alpha \beta  \right. \hspace*{5mm} \non \\
& + & \frac{9}{4} \log x_a \left[ \left( \alpha-\beta + \lam
\right) \left( \lambda -1-\alpha-\beta - \frac{4}{3}\alpha\beta \right)
+(\alpha-\beta)\left( \alpha +\beta+ \frac{20}{9} \right) \right]
- \frac{\pi^2}{4} \alpha  \non \\
& + & \frac{1}{4} \log ^2 x_a \left[ \lambda (2\lambda -2-\alpha-\beta)
-3(\alpha+\beta)-16\alpha \beta+ ( 2\lambda+1+\alpha+\beta) (\alpha-\beta)
\lam\right] \non \\
& + & \left. \frac{3}{4} \log x_a \log x_b (\alpha+\beta)(1-2\alpha \beta)
+\frac{1}{2} \left[4 \alpha \beta-(\alpha+\beta) \lambda \right]
{\cal I} +(1+\alpha+\beta-\lambda ){\cal I}' \right\} \non \\
& + &  \ \ \left\{ \alpha \longleftrightarrow \beta \right\} \non
\en
\be
\Pi_L^- &=& \left\{ - \frac{3}{2\epsilon^2} +\frac{1}{\epsilon} \left( 3\rho_a
-
\frac{11}{4} \right) + \frac{11}{2} \rho_a -\frac{3}{4} (\rho_a+\rho_b)^2
+\frac{3}{8} - 3\alpha - \frac{\pi^2}{4} \right. \non \\
& - & \frac{3}{2} \log x_a \left[ \alpha-\beta + (\alpha+\beta-3)(\alpha-\beta+
\lam ) \right] + \frac{1}{2}\log ^2 x_a \left[ 1+\alpha+\beta-
\lambda +(\beta- \alpha) \lam \right] \non \\
& + & \left. \frac{3}{2} \log x_a \log x_b \left[1+\alpha+\beta-2 \alpha \beta
\right] -(1+\alpha+\beta){\cal I} -2{\cal I}' \right\} \ \ + \ \
\left\{ \alpha \longleftrightarrow \beta \right\}
\en

\vspace*{0.1cm}

\nn with  ${\cal I}$ and ${\cal I}'$ given by
\be
{\cal I} & = & F(1)+F(x_a x_b)-F(x_a)-F(x_b) \non \\
{\cal I}' &= & \lam G(x_a x_b) -\frac{1}{2} (\beta-\alpha+\lam )G(x_a)
-\frac{1}{2}( \alpha - \beta + \lam ) G(x_b)
\en

\vspace*{0.1cm}

\nn In terms of the polylogarithmic functions \cite{levin} ${\rm Li}_2(x)=
-\int_0^1 y^{-1}\log (1-xy) {\rm d}y$ and ${\rm Li}_3(x)= -\int_0^1 y^{-1}
\log y \log(1-xy)  {\rm d}y$, the functions $F$ and $G$ are given by
\be
F(x) & = & 6{\rm Li}_3(x)-4{\rm Li}_2(x) \log x - \log^2 x \log (1-x) \non \\
G(x) &= & 2{\rm Li}_2(x) +2 \log x \log (1-x)+ \frac{x}{1-x} \log^2 x
\en

\nn These two functions  also admit  a simple and useful integral
representation
\cite{bro}
\be
F(x)=\int_0^x dy \left(\frac{\log y}{1-y}\right)^2\log\frac{x}{y}\ \ \ \ ,
\ \ \ \ \ G(x)=\ x \ F'(x)=\int_0^x dy\left(\frac{\log y}{1-y}\right)^2
\en
\vspace*{0.3cm}

\nn The imaginary parts of the vacuum polarization function  are derived along
the same lines as discussed previously in the one--loop case. Using the fact
that
\be
{\cal I}m \log x_{a,b} = \pi \hspace*{0.5cm} , \hspace*{1cm}
{\cal I}m \log^2 x_{a,b}  = 2\pi \log|x_{a,b}| \non
\en
\vspace*{-5mm}
\be
{\cal I}m{\cal I}' &= & \pi {\cal J}' \ = \pi \left\{ \ 4 \lam \left[ \log (1-
x_ax_b)+\frac{x_ax_b}{1-x_ax_b} \log|x_a x_b| \; \right] -(\beta-\alpha+ \lam )
\left[ \log (1-x_a) \right. \right. \non \\
& + & \left. \left. \frac{x_a}{1-x_a} \log |x_a|\right]
- (\alpha-\beta+ \lam ) \left[ \log (1-x_b)+\frac{x_b}{1-x_b} \log |x_b|
\right] \ \right\} \non
\en
\vspace*{-3mm}
\be
{\cal I}m{\cal I}  &= & \pi {\cal J} \ = \ -2 \pi \left[ 4 {\rm
Li}_2 (x_ax_b) -2{\rm Li}_2(x_a)-2{\rm Li}_2(x_b) +2\log|x_a x_b|
\right. \non \\
& \times & \left. \log (1-x_ax_b) -\log|x_a| \log (1-x_a) -
\log|x_b| \log(1-x_b) \right ]
\en

\vspace*{3mm}

\nn one obtains for ${\cal I}m \Pi_{T,L}^\pm$

\be
\frac{1}{\pi} {\cal I}m \Pi_T^+(s) &=& \frac{1}{6}(11+19\alpha+19\beta
+12\alpha
\beta- 5 \lambda ) \lam +\frac{4}{3}(\alpha-\beta) \left[\log (1-x_b) -\log
(1-x_a) \right] \non \\
&+ & \frac{1}{6} \left[ 3(1+\alpha+\beta) ((\beta-\alpha) \lam +
1+\alpha+\beta+2\alpha \beta- \lambda )+8\alpha+26 \alpha \beta
\right] \log |x_a| \non \\
&+ & \frac{1}{6} \left[ 3(1+\alpha+\beta) ((\alpha-\beta) \lam +
1+\alpha+\beta+2\alpha \beta- \lambda )+8\beta+26 \alpha \beta
\right] \log |x_b| \non \\
&+ & \frac{1}{3} \left[(\alpha+\beta-2) \lambda -12 \alpha \beta \right]{\cal
J}
 - \frac{2}{3} \left[ 3(1+\alpha+\beta)- \lambda \right] {\cal J}' \non \\
& & \non \\
\frac{1}{\pi} {\cal I}m \Pi_T^- (s) & =& (9+\alpha+\beta) \lam
+\left[4(1+\alpha+
\beta) +2\alpha \beta- \lambda +(\beta-\alpha) \lam \right]\log |x_a| \non \\
&+ & \left[4(1+\alpha+\beta)+2\alpha \beta- \lambda +(\alpha-\beta) \lam
\right]\log |x_b|-2(1+\alpha+\beta){\cal J} -4{\cal J}' \non \\
& & \non \\
\frac{1}{\pi} {\cal I}m \Pi_L^+ (s) &=& \frac{1}{2}(9 \lambda -9-9\alpha-9\beta
-12\alpha \beta) \lam -[(\alpha+\beta) \lambda -4\alpha\beta]{\cal J}
+2[1+\alpha+\beta- \lambda ]{\cal J}' \non \\
&+ & \frac{1}{2} \left[ (2 \lambda +1+\alpha+\beta) \lam (\alpha-\beta +
\lam )-( \lambda+3\alpha\beta)(2\alpha+2\beta+3)-7\alpha\beta \right]
\log |x_a| \non \\
&+ & \frac{1}{2} \left[ (2 \lambda +1+\alpha+\beta) \lam (\beta-\alpha+
\lam )-(\lambda +3\alpha\beta)(2\alpha+2\beta+3)-7\alpha\beta \right]
\log |x_b| \non \\
& & \non \\
\frac{1}{\pi} {\cal I}m \Pi_L^- (s) &=& 3(3-\alpha-\beta) \lam
+\left[4(1+\alpha+\beta)
-6\alpha \beta-\lambda +(\beta-\alpha) \lam \right]\log |x_a| \non \\
&+ & \left[4(1+\alpha+\beta)-6\alpha \beta- \lambda +(\alpha-\beta) \lam
\right]\log |x_b| -2(1+\alpha+\beta){\cal J} -4{\cal J}'
\en

\vspace*{4mm}

\nn These expressions for the real and imaginary parts are the main result of
this paper. \\

\nn Close forms for ${\cal I}m \Pi_{T,L}^{V,A}(s)$ in the general case $m_a
\neq m_b \neq 0$ have been also derived in the past by a number of authors
\cite{cgv,rry,stn} [in the first reference only the transverse part is
given] by directly calculating the QCD corrections to the flavor changing decay
of a vector boson. The results that we obtain here using a completely different
method, agree with those of Ref.~\cite{cgv} and also with Ref.~\cite{stn} once
some obvious mistakes in the integrals of their Appendix [$J_1$ and $J_2$] are
corrected; see also Ref.~\cite{den}.

\newpage

\subsection*{5.~Special Cases}
\renewcommand{\theequation}{5.\arabic{equation}}
\setcounter{equation}{0}

\nn From the general formulae given in the previous section one can derive the
expressions of the vector and axial--vector components of the self--energies
in various special situations of physical relevance. Here we will exhibit these
expressions in four different cases: $i)$ the two quarks have equal masses,
$ii)$ one of the quarks is massless\footnote{In fact, we have expanded the
expressions of the previous sections around $m_b=m_a$ and $m_b=0$ retaining
terms up to order $(m_b-m_a)^2/s$ and $m_b^2/s$, respectively. The rather
lengthy expressions can be found in Ref.~\cite{paolo}.} and $iii)$ the quark
masses are much larger or much smaller than the momentum transfer. The results
that we obtain in these special cases provide several checks of the general
expressions of the previous section and allow for a comparison with various
results available in the literature. \\

\nn $i)$ Case $m_b=m_a$ \s

\nn  The real parts are given by
\be
\Pi_T^{V}(s) &=& s \left\{ \frac{1}{2 \epsilon} -\rho +\frac{55}{12}-
\frac{26}{3}\alpha +\sqrt{1+4\alpha} (1-6\alpha) \log x -\frac{2}{3} \alpha
(4+\alpha) \log ^2 x \right. \non \\
&+& \left. \frac{2}{3}(4\alpha^2-1) \left[ F(1)+F(x^2)-2F(x) \right] -
\frac{4}{3} (1-2\alpha) \sqrt{1+4\alpha} \left[ G(x^2)-G(x) \right] \right\}
\non \\
\Pi_T^{A}(s) &=& s \left\{ - \frac{6\alpha}{\epsilon^2} + (1+24 \alpha \rho
-22\alpha) \frac{1}{2 \epsilon} - (1+ 12 \alpha \rho -22\alpha) \rho +
\frac{55}{12}- \frac{19}{6} \alpha +4 \alpha^2 - \alpha \pi^2 \right. \non \\
&+& (1+12\alpha +4\alpha^2) \sqrt{1+4\alpha} \log x + \frac{2}{3} \alpha
(5+11\alpha+6\alpha^2 )\log ^2 x -\frac{2}{3}(1+2\alpha) \non \\
& \times & \left. (1+4\alpha) \left [F(1)+F(x^2)-2F(x) \right]
- \frac{4}{3} (1+4\alpha)^{3/2} \left[ G(x^2)-G(x) \right]
\right\} \non \\
\Pi_L^{A}(s) &=& 2s \alpha \left\{ -\frac{3}{\epsilon^2} +\left( 6\rho
-\frac{11}{2} \right) \frac{1}{\epsilon} +\frac{3}{4} -6 \alpha - \frac{1}{2}
\pi^2 + 3(3-2\alpha) \sqrt{1+4\alpha} \log x \right. \non \\
&+& (11-6\rho) \rho + (3+4 \alpha-6 \alpha^2) \log^2 x - 2 (1+2\alpha) \left[
F(1)+F (x^2)-2F(x) \right] \non \\
&-& \left. 4 \sqrt{1+4\alpha} \left[ G(x^2)-G(x) \right] \right\}
\en

\nn with $x = 4\alpha / (1+\sqrt{1+4\alpha})^2$, and the imaginary parts are
\be
{\cal I}m \Pi_T^{V}(s) &=& \frac{2\pi}{3} s\left\{ \frac{3}{2} \lam
(1-6\alpha)-2 \alpha(4+\alpha) \log|x| - (1-4\alpha^2) J -2(1-2\alpha) J'
\right\} \non \\
{\cal I} m\Pi_T^{A}(s) &=& \frac{2\pi}{3} s\left\{ \frac{3}{2}
\lam (1+12 \alpha +4 \alpha^2 )+2\alpha(5+11\alpha+6\alpha^2 )\log|x|-(1+2
\alpha) \lambda J-2\lambda J' \right\}\non \\
{\cal I}m\Pi_L^{A} (s) &=& 4\pi s\alpha \left\{ \frac{3}{2}\lam (3-2\alpha)
+(3+4\alpha-6\alpha^2) \log|x|-(1+2\alpha)  J - 2J'\right\}
\en
\nn where
\vspace*{-2mm}
\be
J &= & -4 \left[ {\rm Li}_2 (x^2) -2{\rm Li}_2(x)
+2\log|x| \log (1-x^2) - \log|x| \log (1-x) \right] \non \\
J' &= & 2 \sqrt{1+4\alpha} \left[ 2 \log (1-x^2) - \log (1-x)
+\frac{x(3x -1)}{1-x^2} \log|x| \right]
\en

\nn The vector part of the longitudinal component vanishes in this case: this
is expected to occur as a consequence of the QED--like Ward identity. \s

\nn In the equal mass case, $\Pi_{T}^{V}(s)$ coincides, up to a color
factor, with the irreducible part of the photon self--energy in QED which has
been calculated in the fifties by the pioneers of Ref.~\cite{QED}.
$\Pi_{T}^{V,A}(s)$ has also been derived using a dispersive approach
in \cite{cgv}, and more recently in \cite{kn}. \s

\nn $ii)$ Case $m_b=0$ \s

\nn In this case,  the coefficients of $\Pi^-_{L,T}(s)$ vanish so that $\Pi_{L,
T}^V (s)= \Pi_{L,T}^A (s)= s\Pi^{+}_{L,T}(s)$, and one obtains for the real
parts
\be
\Pi_T^{V,A}(s) &=& s \left\{ -{3\alpha\over2\epsilon^2} + \left(1-{11
\over 2} \alpha+ 6 \alpha\rho \right) {1\over 2\epsilon}- \left( 1-\frac{11}{2}
\alpha +3\alpha \rho \right) \rho +{55\over12}-{71\over24}\alpha \right. \non
\\
&-& \frac{\pi^2}{4} \alpha -
{5\over6}\alpha^2+(1+\alpha)\left(1+{3\over2}\alpha-
{5\over6}\alpha^2
\right)\log x +{\alpha\over6}(1-3\alpha)(1+\alpha)\log^2 x \non\\
&+& \left. {1\over3}(2+\alpha)(\alpha-1)G(x) +{1\over3}(\alpha-2)(1+\alpha)^2
\left[ F(1)-F(x) \right] \right\} \non \\
\Pi_L^{V,A}(s) &=& s \alpha \left\{ -{3\over2\epsilon^2} + \left(-{11
\over4}+ 3\rho \right){1\over\epsilon} +{3\over8} + {7\over2}\alpha +{11\over2}
\rho -3\rho^2 - \frac{\pi^2}{4} +{9\over2}(1+\alpha)^2\log x \right. \non\\
&+& \left. {1\over2}(3+2\alpha)(1+\alpha)^2\log^2 x + (1+\alpha) G(x)
-(1+\alpha)^2 \left[ F(1)-F(x) \right] \right\}
\en

\nn with $x=\alpha/(1+\alpha)$, and for the imaginary parts
\be
{\cal I}m\Pi_T^{V,A}(s) & = & \pi s \left\{ 1+{5\over2}\alpha +{2\over3}
\alpha^2 -{5\over6}\alpha^3 -{1\over3}(1+\alpha)^2(4-5\alpha)\log(1+\alpha)
-{\alpha\over3}(5\alpha^2+4\alpha-5) \right. \non \\
& \times & \left. \log (-\alpha) - \frac{2}{3} (2-\alpha) (1+\alpha)^2 \left[
2{\rm Li}_2 \left(\alpha\over \alpha +1 \right)- \log (1+\alpha) \log{-\alpha
\over 1+\alpha} \right] \right\} \non \\
{\cal I}m\Pi_L^{V,A} (s) & = & - \pi s \alpha(1+\alpha) \left\{(1+\alpha)
 \left[ -\frac{9} {2}+ (5+2\alpha) \log(1+\alpha) \right] -(3+7\alpha+2\alpha^2
) \log(-\alpha) \right. \non \\
&+& \left. 2(1+\alpha) \left[ 2{\rm Li}_2 \left(\alpha\over \alpha +1 \right)
- \log (1+\alpha) \log{-\alpha \over 1+\alpha} \right] \right\}
\en

\nn We see that in this limit, the self--energies are free of mass
singularities
as required by the Kinoshita--Lee--Nauenberg theorem \cite{kln}. The expression
of $\Pi_{T}^{V,A}(s)$ in this special case has been obtained in Ref.~\cite{kn}
by means of a dispersion integration. \s

\nn Note that the results for the real parts of the vector bosons vacuum
polarization function in these two special cases slightly differ from those
obtained some time ago by one of the present authors in \cite{moi2}: the
$\pi^2$ terms are absent in the expressions given in the latter reference [this
is due to the fact that the $(1+ \epsilon^2 \pi^2/12)$ factor in the mass
counterterm eq.~(3.5) has been omitted\footnote{We thank B. Kniehl for bringing
this point to our attention.}]. However, this discrepancy does not
affect the radiative corrections to physical quantities when evaluated in the
on--shell scheme, which is commonly used to calculate radiative corrections to
electroweak observables. Indeed, these $\pi^2$ terms are related to the $1/
\epsilon^2$ poles which must cancel in physical quantities and therefore they
do not appear in the radiative corrections to observables like the $\rho$
parameter, the $W$ boson mass or the effective weak mixing angle $\sin^2
\theta_W$. \s

\nn $iii)$ Cases $s \gg m_{a,b}^2$ and $s \ll m_{a,b}^2$  \s

\nn In the limit of zero momentum transfer, the expressions of $\Pi^{V,A}_{L,T}
(s)$ simplify considerably, and one obtains [we use the fact that  $G(x)+G(1/x)
=2\pi^2/3$]
\be
\Pi_{T,L}^{V,A} (0) & = & \frac{3}{2\epsilon^2} (m_a^2+m_b^2)+
\frac{1} {\epsilon} \left[ \frac{11}{4}(m_a^2+m_b^2)-3m_a^2 \rho_a-3m_b^2
\rho_b \right]  - \frac{11}{2} (m_a^2 \rho_a+ m_b^2\rho_b) \non \\
&+ &  3m_a^2 \rho_a^2 +3m_b^2 \rho_b^2 + \frac{35}{8}(m_a^2+m_b^2) +\frac{1}{4}
(m_a^2-m_b^2) \left [G\left(\frac{m_a^2}{m_b^2} \right) -G\left( \frac{m_b^2}
{m_a^2} \right) \right] \non \\
&+& (m_a^2+m_b^2) \frac{\pi^2}{4} + \frac{m_a^2m_b^2}{m_a^2-m_b^2} \log
\frac{m_a^2}{m_b^2} + m_a^2 m_b^2 \frac{m_a^2+m_b^2}{(m_a^2-m_b^2)^2} \log^2
\frac{m_a^2}{m_b^2} \non \\
&\pm  & \ m_a m_b \ \left[ -\frac{3}{\epsilon^2}+ \frac{1}{\epsilon} \left(
3\rho_a + 3\rho_b-\frac{11}{2} \right) -\frac{3}{2} (\rho_a + \rho_b)^2 +
4 (\rho_a + \rho_b) -\frac{31}{4} \right. \non \\
&+ &  \left. 3 \frac{m_a^2 \rho_b- m_b^2\rho_a}
{m_a^2-m_b^2}+3\frac{m_a^2m_b^2}
{(m_a^2-m_b^2)^2} \log^2 \frac {m_a^2}{m_b^2} - \frac{1}{2} \pi^2 \ \right]
\en

\nn As expected, the self--energies are regular in this limit and one has
again $\Pi_{T}^{V,A}(0)=\Pi_{L}^{V,A}(0)$. In the two special cases $m_a=m_b$
and $m_b=0$ this expression simplifies to [$\Pi_{V}(0)$ vanishes in the equal
mass case]
\be
\Pi_{T,L}^{A}(0,m_a=m_b) &=& m_a^2 \left[ \frac{6}{\epsilon^2} +\frac{1}{
\epsilon} (11-12 \rho_a) +12 \rho_a^2 - 22 \rho_a + \frac{31}{2} + \pi^2
\right]  \non \\
\Pi_{T,L}^{V,A}(0,m_b=0) &=& \frac{m_a^2}{4} \left[\frac{6}{\epsilon^2}
+\frac{1}{\epsilon} (11-12 \rho_a) +12 \rho_a^2 - 22 \rho_a + \frac{35}{2}
+ \frac{5}{3} \pi^2 \right]
\en

\nn Note also the presence of the extra $\pi^2$ terms compared to
Ref.~\cite{moi2} as discussed previously. \s

\nn In the opposite limit $ |s|\gg m_{a,b}^2$, the vector and axial--vector
components are simply
\be
\Pi_T^{V,A} (s) &\sim & s \left[ \frac{1}{2\epsilon} - \log \frac{-s}
{\mu^2} +\frac{55}{12} - 4 \zeta (3) \right] \non \\
\Pi_L^{V,A} (s) &\sim & 0
\en

\nn with $\zeta(3)=F(1)/6=1.202$, in agreement with Ref.~\cite{fnr}. Here
again, there is no mass singularity and the quadratically divergent components
$\Pi_{T}^{V,A}(s)$ obey the chiral symmetry relation. The next term of this
expansion, i.e. the terms of ${\cal O}(m_{a,b}^2/s)$, can be found in
Ref.~\cite{nous}. \\

\nn In view of the several checks that we have performed and of the various
results available in the literature that we recover in the special situations
discussed above, we have good confidence that the expressions that we give
here for the general case $m_a \neq m_b \neq 0$ are correct.

\newpage

\subsection*{6. Summary}

In this paper, the contribution of heavy quarks to the vacuum polarization
function of the electroweak gauge bosons, and to the mixing amplitude of two
gauge bosons, were calculated at first order in the strong interaction. Both
the transverse and longitudinal components of the vacuum polarizations
functions
were given in the most general case: arbitrary masses for the quarks and finite
momentum transfer. Full analytical formulae for the real as well as for the
imaginary parts at ${\cal O} (\alpha \alpha_S)$ were presented in the on--shell
quark mass scheme. They are given in a compact form in terms of polylogarithmic
functions which can be easily evaluated using standard computer routines. \s

\nn The dependence of the result on the definition of the quark masses has been
discussed in detail. We provided the necessary expressions which allow, having
at hand the result in the on--shell mass scheme, to obtain the vector bosons
self--energies for quark masses defined in any other renormalization scheme. \s

\nn The expressions of the self--energies were then derived in two situations
of great phenomenological interest: first the case where the two quarks have
equal masses which corresponds to the photon and the $Z$ boson self--energies
and second, the case where one of the quarks has a negligible
mass with respect to the other which corresponds to the approximate
contribution of the top--bottom isodoublet to the charged W boson self--energy.
The expressions of the self--energies for asymptotic values of the momentum
transfer: $s \to \infty$ or $s \to 0$ were also given. \s

\nn Some of the results in the special cases that we have considered here
for the real parts of the vacuum polarization functions, as well as the
imaginary parts in the general case, are available in the literature and the
comparison with them provided very powerful checks of our calculation in the
general case. \s

\vspace*{1cm}

\nn {\bf Acknowledgements.} \s

\nn Discussions with Bernd Kniehl and Alberto Sirlin are gratefully
acknowledged. One of us (PG) would like to thank the Physics Department
of the University of Torino for its warm hospitality during this summer.
This work is partially supported by the National Sciences and Engineering
Research Council of Canada and by the National Science Foundation under
Grant No. PHY--9017585.

\newpage

\subsection*{APPENDIX: Scalar Integrals}

\renewcommand{\theequation}{A.\arabic{equation}}
\setcounter{equation}{0}

In this appendix we will give for completeness the expressions of the one and
two--loop scalar integrals which are needed in the evaluation of the two--loop
vacuum polarization functions \cite{bro,gen,moi2}. We will closely follow the
notation of Ref.~\cite{moi2}, but simplify further some of the expressions and
correct a few misprints which occured in the Appendix. \s

\nn {\bf One--loop integrals.} \s

\nn These are needed in the calculation of the amplitude both at the one--loop
and the two--loop levels. Indeed, when expressing the
scalar products of momenta appearing in the numerator of eq.~(3.1) in terms of
combinations of the polynomials in the denominators, the gluon propagator [i.e.
the $(k_1-k_2)^2$ factor] cancel in some cases and the two--loop amplitude
reduce to the product of two one--loop integrals. These one--loop intregrals
are
\be
D_{A} &=& - \frac{i}{q^2} (-q^2 e^\gamma)^\epsilon \int \frac{d^nk}{\pi^{n/2}}
\ \frac{1}{k^2-m_{a}^2 } \non \\
&=& - \alpha \left[ \frac{1}{\epsilon} +1- \log \alpha +\epsilon \left(
\frac{1}{2} \log^2 \alpha - \log \alpha +1 + \frac{\pi^2}{12} \right) \right]
\en
[and similarly for $D_{B}$ where one has to replace $\alpha =-m_a^2/q^2$ by
$\beta=-m_b^2/q^2$] and using the variables $x_{a,b}, \lambda$ defined in
eqs.~(2.9) and (2.10)
\be
K &=& - i (-q^2 e^\gamma)^\epsilon \int \frac{d^nk}{\pi^{n/2}}
\ \frac{1}{[k^2 -m_a^2][(k-q)^2-m_b^2] } \non \\
&=& \frac{1}{2} \left[ \frac{1}{\epsilon} + 2+ (\alpha - \beta + \lam) \log x_a
- \log \alpha \right] + \frac{1}{2} \left[ \alpha \leftrightarrow \beta
\right] + {\cal O} (\epsilon)
\en
Note that in the expression of $K$, the ${\cal O}(\epsilon)$ term is not
needed: its total contribution, via terms likes $K^2,~K D_{A},~K \partial K/
\partial \alpha, \ \cdots$ , vanishes in the final result. \s

\nn The derivatives of $K$ and $D_A$ with respect to $\alpha$ will enter the
expressions of the two--loop integrals and are given by
\be
\frac{\partial D_A}{\partial \alpha} &=& - \frac{1}{\epsilon} \left[ 1-
\epsilon \log \alpha +\epsilon^2 \left( \frac{1}{2} \log^2 \alpha +
\frac{\pi^2}{12} \right) \right]  \non \\
\frac{\partial K}{\partial \alpha} &=& \frac{1}{2 \lam} \left[ (1+\alpha-\beta
+\lam ) \log x_a + (1+\alpha -\beta - \lam ) \log x_b \right]
\en

\vspace*{0.3cm}

\nn {\bf Two--loop integrals.} \s

\nn To simplify the notation, we shall use for the definition of the two--loop
integrals the following abreviations:
\be
<X> = - (-q^2 e^\gamma)^{2\epsilon} \ \int \frac{d^nk_1 }{\pi^{n/2}}
\ \int \frac{d^nk_2}{\pi^{n/2}} \ \ X
\en
and
\be
K_{1,2} \equiv k_{1,2}^2-m_a^2 \ \ \ , \hspace*{0.9cm}
K_{0} \equiv (k_1-k_2)^2 \ \ \ , \hspace*{0.9cm}
Q_{1,2} \equiv (k_{1,2}-q)^2-m_b^2 \ \ \
\en

\nn With the help of the variables defined in eqs.~(2.9)--(2.10), the two--loop
integrals are
\be
q^2 P & \equiv  & < \frac{K_0}{K_1 K_2  Q_1 Q_2 } > =
(1-\alpha+\beta) KD_A + (1+\alpha-\beta) KD_B - \frac{1}{2} (D_A-D_B)^2
- \frac{\lambda}{2}  K^2 \non \\
W_A & \equiv & \frac{1}{(q^2)} < \frac{1}{K_1^2K_2} > = -D_A
\frac{\partial D_A}{\partial \alpha} \non \\
T_A & \equiv   & \frac{1}{(q^2)^2} < \frac{Q_1}{K_1^2K_2} > = D_A^2-D_A
(1-\alpha +\beta ) \frac{\partial D_A}{\partial \alpha} \non \\
U_A & \equiv & < \frac{1}{K_1^2K_2Q_1} > = -D_A \frac{\partial K}{\partial
\alpha} \non \\
M_A & \equiv & \frac{1}{q^2} < \frac{1}{K_1K_2K_0} > = - \frac{1}{\alpha}
\frac{1- \epsilon}{1-2\epsilon} D_A^2 \non \\
V_A & \equiv & < \frac{1}{K_1^2 K_2 K_0} > = - \frac{1}{2\alpha^2} \frac{1-
\epsilon} {1-2\epsilon} D_A \left( D_A -2 \alpha \frac{\partial D_A}{\partial
\alpha} \right) \non \\
R_A & \equiv & \frac{1}{q^2} < \frac{Q_1}{K_1^2 K_2 K_0} > = M_A + (1- \alpha +
\beta ) V_A \non \\
N_A & \equiv & \frac{1}{q^2} < \frac{(k_1-2k_2)\cdot q}{K_1 K_2 K_0 Q_1} > =
\alpha \left[ 1+ \frac{ \log x_a }{1-x_a} \right] \left[ 1 +\frac{x_b \log x_b}
{1-x_b} \right] \non \\
J_A & \equiv  & < \frac{1}{K_1 K_2 K_0 Q_1} > = \frac{1}{2}(1+\epsilon)K^2 +
\frac{1}{2} \left[ 3- \frac{x_a \log^2 x_a}{(1-x_a)^2}-\frac{x_b \log^2 x_b}
{(1-x_b)^2} - G(x_a)+ G(x_b) \right] \non \\
q^2 L & \equiv  & < \frac{1}{K_1 Q_2 K_0 } > = \frac{1}{2}\left[ \frac{7}
{8}+ \frac{1}{4} (2+\epsilon)K^2 + M_A+N_A-(1-\alpha+\beta)J_A \right] +
\frac{1}{2} [\alpha \leftrightarrow \beta] \non \\
Q_A & \equiv & \frac{1}{q^2} < \frac{Q_1}{K_1 K_2 Q_2 K_0 } > = \frac{1}{2}
\left[ (1-\alpha+\beta)J_A -L +M_A \right] + N_A \non \\
S_A & \equiv   & < \frac{q^2}{K_1^2 K_2 Q_1 K_0 } > =  \frac{1}{\lam} \left[
\frac{\log x_a}{1-x_a} \left( 1+\frac{\log x_a}{1-x_a} \right) - \frac{x_b
\log x_b}{1-x_b} \left( 1+\frac{x_b\log x_b}{1-x_b} \right) \right]
- (1+\epsilon) \non \\
& & \hspace*{0.2cm} \times K \frac{\partial K}{\partial \alpha}  + \frac{1}{2
\alpha} \left[ \lam G(x_a x_b) -\frac{1}{2}
(1-\alpha +\beta + \lam) G(x_a) +\frac{1}{2} (1-\alpha +\beta  - \lam) G(x_b)
\right] \non \\
I & \equiv  & < \frac{q^2}{K_1 K_2 Q_1 Q_2 K_0 } > =  F(1) + F(x_a x_b)-
F(x_a) - F(x_b)
\en
The expressions of the functions $F$ and $G$ are given in eqs.~(4.4) and (4.5).
\s

\nn All the other two--loop integrals can be straightforwardly obtained by
taking advantage of the symmetry in the interchange of the integration
variables $k_1$ and $k_2$ as well as in the interchange of the masses $m_a$
and $m_b$.

\vspace*{2cm}

\subsection*{Figure Captions}

\vspace*{0.5cm}

\begin{description}
\item[Fig.~1] Feynman diagrams for the contribution of quark pairs to the
vacuum polarization function of a gauge boson at the one--loop (a) and
two--loop (b) levels. \\

\item[Fig.~2] Feynman diagrams for the one--loop quark self--energy diagram (a)
and for the contribution of the mass counterterm to the polarization function
at the two--loop level (b).

\end{description}

\end{document}